\documentclass[american,aps,pra,reprint,groupedaddress,twocolumn,showpacs]{revtex4-1} 
\usepackage[unicode=true,pdfusetitle, bookmarks=true,bookmarksnumbered=false,bookmarksopen=false, breaklinks=false,pdfborder={0 0 0},backref=false,colorlinks=false] {hyperref}
\hypersetup{ colorlinks,linkcolor=myurlcolor,citecolor=myurlcolor,urlcolor=myurlcolor}
\usepackage{braket,colortbl,amsthm,amsmath,cleveref,amssymb,txfonts}
\definecolor{myurlcolor}{rgb}{0,0,0.7}
\usepackage{graphics,graphicx}
\usepackage{color}
\usepackage{colortbl}
\usepackage{subfigure}
\usepackage{graphicx}
\usepackage[font=small,labelfont=bf,
   justification= centerlast,
   format=hang]{caption}
\usepackage{natbib}
\usepackage{ragged2e}
\usepackage{amsmath}
\usepackage{graphicx}
\usepackage[utf8x]{inputenc}
\usepackage{color}
\usepackage{amsmath}
\usepackage{tikz}
\usetikzlibrary{shapes.geometric, arrows}
\usepackage{latexsym}
\usepackage{amssymb}
\usepackage{amsthm}
\usepackage{bm}
\usepackage{graphics,epstopdf}
\usepackage{color}
\usepackage{amsmath}

\usepackage[usenames,dvipsnames,svgnames]{pstricks}
\usepackage{epsfig}
\usepackage{pst-grad} 
\usepackage{pst-plot} 
\tikzstyle{startstop} = [rectangle, rounded corners, minimum width=3cm, minimum height=1cm,text centered, draw=black, fill=red!30]

\tikzstyle{env}=[circle,  ball color = green!20, minimum size= 80mm]
\tikzstyle{central}=[circle, ball color = red!100, minimum size=8mm]
\tikzstyle{bath}=[circle, ball color =blue!75, minimum size=4mm]

\theoremstyle{plain}

\def\bea{\begin{eqnarray}}
\def\eea{\end{eqnarray}}
\def\ba{\begin{array}}
\def\ea{\end{array}}

\def\beq{\begin{equation}}
\def\eeq{\end{equation}}

\begin{document}


\title{Witnessing nonseparability of bipartite quantum operations}

\author{Sohail}
\email{sohail@hri.res.in}
\affiliation{Quantum Information and Computation Group, Harish-Chandra Research Institute, HBNI, Jhunsi, Allahabad 211 019, India}

\author{Ujjwal Sen}
\email{ujjwal@hri.res.in}
\affiliation{Quantum Information and Computation Group, Harish-Chandra Research Institute, HBNI, Jhunsi, Allahabad 211 019, India}

\begin{abstract}
We provide a method for witnessing nonseparability of quantum processes on shared systems, which uses the channel-state duality. The method uses a maximally entangled state as a resource. We show that using the resource provides significant advantage over the corresponding protocol for nonseparability detection without the resource.

\end{abstract}

\maketitle
\section{Introduction}
Creation and manipulation of entanglement \cite{RevModPhys.81.865} are one of the basic necessities of quantum information tasks. Local quantum operations and classical communication between parts of a physical system cannot create entangled states between the parts. It is therefore important to identify quantum processes on shared systems that are not local. An efficient method for detecting entanglement of shared quantum states is by using entanglement witnesses \cite{woronowicz1976,CHOI1984462,1996PhLA..223....1H,1998quant.ph.10091T,PhysRevA.62.052310,2002JMOp...49.1399B,PhysRevA.66.062305,2003JMOp...50.1079G,PhysRevLett.92.087902,das2017separability}. We show that a similar technique can be utilized for witnessing nonseparability of quantum processes on shared systems.
\\
\\
Just like the concept of entanglement witnesses for states, the technique of nonseparability witnesses for quantum operations also uses the mathematics around the Hahn-Banach theorem of normed linear spaces \cite{smith_1970, loTa-kambal}. Additionally, we use the channel-state duality between quantum states and quantum operations \cite{CHOI1975285,article, kraus1983states,Sudarshan1985}. The proposed method requires a maximally entangled state as a resource.
\\
\\
We exemplify our method by considering the paradigmatic controlled-NOT (CNOT), square-root of swap, and Bell gates on two-qubit systems. We also analyze the noise threshold that our method can withstand by considering  noisy versions of the  gates, where the noise is modelled by the completely depolarizing channel. This also helps us to identify a trade-off between the noise threshold and the entanglement in the resource.
\\
\\
We begin in Section \ref{eibar-bachhadhan-chaT-bogole} with a short discussion on the Hahn-Banach theorem. This is followed by a short discussion on the channel-state duality in Section \ref{saRe-chuattar}. The results are presented in Section \ref{preme-paRa-baran}, followed by a conclusion in Section \ref{karane-akaran}.
\section{Gathering the tools}
\subsection{The Hahn-Banach theorem}
\label{eibar-bachhadhan-chaT-bogole}
The Hahn-Banach theorem \cite{smith_1970, loTa-kambal} is a very important and powerful tool in functional analysis. It is an important contraption in finding whether a given quantum mechanical state of a multiparty physical system is entangled. Instead of the usual version, it is a corollary (Corollary 2 below, which has been referred to as the ``hyperplane separation theorem'' \cite{loTa-kambal}) of the theorem that is utilized for the purpose.
\\
\\
\textbf{Hahn-Banach theorem:}  Let $M$ be a linear subspace of a normed linear space $N$, and let $f$ be a functional defined on $M$. Then $f$ can be extended to a functional $f_0$ defined on the whole space $N$ such that 
$ \left\Vert f_0 \right\Vert = \left\Vert f \right\Vert $.
\\
\\
The following corollaries are consequences of the theorem.
\\
\\
\textbf{Corollary 1}:  If $M$ is a closed linear subspace of a normed linear space $N$ and $x_0$ is a vector not in $M$, then there exists a functional $f^0$ in $N^*$ such that $f^0 (M)=0$ and $f^0 (x_0)\neq 0$, where $N^*$ is the dual space of $N$.
\\
\\
\textbf{Corollary 2}: Let $M$ be a convex compact set in a finite dimensional Banach space $X$. Let $\rho \notin M $ be a point in $X$. Then there exists a hyperplane that separates $\rho$ from $M$. 
\\
\\
\textbf{Entanglement witness and Hahn-Banach theorem:}
It is known that the set of separable states of a bipartite quantum system is a convex compact subset of the set of quantum states  \citep{PhysRevA.40.4277}. So it is clear from corollary 2 that there exists a functional in the dual space of the Hilbert space of the physical system, which can distinguish  an entangled state from the separable states. It can be constructed as a linear functional on $\mathcal{B}(H)$ where $H$ is the joint Hilbert space  of the bipartite  system, in such a way that it assigns positive real numbers to the separable states and a negative real number to the given entangled state. The linear functional is of the form tr$(W \cdot)$, where $W$ is an operator on $H$, being called an ``witness operator". It has been 
shown  \cite{das2017separability} that $W$ can be chosen as the partial transpose of the projector in the direction of an eigenvector of the given entangled state corresponding to a negative eigenvalue, provided the given entangled state does have a negative eigenvalue after partial transposition. 
\\
\subsection{The CJKS Isomorphism}
\label{saRe-chuattar}
Let $H_1={\mathbb{C}^n}$ and let $\phi : \mathcal{B}(H_1)\rightarrow \mathcal{B}(H_2)$ be a linear map, where $H_1$ and $H_2$ are two Hilbert spaces. Let $ \{e_{ij}\}$, $i,j=1,2,....,n$ be a complete set of matrix units for $\mathcal{B}(H_1)$. Then the Choi-Jamio{\l}kowski-Kraus-Sudarshan (CJKS) matrix \cite{book,article,CHOI1975285,kraus1983states,Sudarshan1985} for $\phi$ is defined to be the operator
                      $\rho_\phi = $ $\sum_{{i,j}=1}^{n} $ $e_{ij} \otimes \phi(e_{ij})$ $\epsilon $ $\mathcal{B}(H_1)\otimes \mathcal{B}(H_2) $.
                      \\
 \\The map $\phi \rightarrow \rho_\phi $ is linear and bijective. This map is called the CJKS isomorphism. Using this isomorphism, the concept of ``channel-state duality" emerges. Here, a channel or quantum channel is a completely positive trace-preserving map, which acts on the space of bounded operators on a Hilbert space. To understand the ``channel-state duality" we need to have a look at the CJKS theorem on completely positive maps.
\\
\\
\textbf{CJKS theorem on completely positive maps:} The CJKS matrix, $\rho_\phi = $ $\sum_{{i,j}=1}^{n} $ $e_{ij} \otimes \phi(e_{ij})$ $\epsilon $ $\mathcal{B}(H_1)\otimes \mathcal{B}(H_2) $, is positive if and only if the map $\phi : \mathcal{B}(H_1)\rightarrow \mathcal{B}(H_2)$ is completely positive.
\\ 
\\
The CJKS isomorphism, with the help of the CJKS theorem on completely positive maps, allows us to view completely  positive trace-preserving linear maps acting on quantum states as a quantum state in a higher-dimensional Hilbert space. If we consider quantum states which are density matrices on an $n$-dimensional Hilbert space, then the completely positive trace-preserving map acting on them can be identified with a density matrix on an  $n^2$-dimensional Hilbert space.
\\
\\
\textbf{Lemma:} The CJKS isomorphism is continuous.
\\
\\
\textbf{Proof:} Let $\phi  \epsilon \mathcal{L}(\mathcal{B}(H_1),\mathcal{B}(H_2))$, where $\mathcal{L}(\mathcal{B}(H_1),\mathcal{B}(H_2))$ is the set of all linear maps from $\mathcal{B}(H_1)$ to $\mathcal{B}(H_2)$, for two Hilbert spaces $H_1$ and $H_2$. The map  $f:\mathcal{L}(\mathcal{B}(H_1),\mathcal{B}(H_2)) \rightarrow \mathcal{B}(H_1)\otimes \mathcal{B}(H_2) $, defined by
$f(\phi)=\rho_\phi=\sum_{{i,j}=1}^{n} $ $e_{ij} \otimes \phi(e_{ij})$, is the CJKS isomorphism. Let ${\phi_n}$ be a sequence in $\mathcal{L}(\mathcal{B}(H_1),\mathcal{B}(H_2))$ which converges to $\phi$ in $\mathcal{L}(\mathcal{B}(H_1),\mathcal{B}(H_2))$. Then 
\(\left\Vert \rho_{\phi_n}-\rho_{\phi} \right\Vert\)
\(=\left\Vert \sum_{{i,j}=1}^{n}  e_{ij} \otimes \phi_n(e_{ij})-\sum_{{i,j}=1}^{n}  e_{ij} \otimes \phi(e_{ij})  \right\Vert\)
\(=\left\Vert\sum_{{i,j}=1}^{n}  e_{ij} \otimes (\phi_n-\phi)(e_{ij}) \right\Vert
\)
\(\leq \sum_{{i,j}=1}^{n} \left\Vert e_{ij} \right\Vert \left \Vert (\phi_n-\phi)(e_{ij}) \right\Vert\)
\(\leq \left(\sum_{{i,j}=1}^{n} \left\Vert e_{ij} \right\Vert \right)^2 \left \Vert (\phi_n-\phi)\right\Vert\).  
  So, the sequence $\rho_{\phi_n}$ converges to  $\rho_\phi$. This implies that the map $f$, the CJKS isomorphism, is continuous. \hfill \(\square\)

\section{Nonseparability witness for quantum operations}
\label{preme-paRa-baran}
We will in this section provide a general method for witnessing nonseparability of bipartite quantum operations.
\\
\\
\textbf{Local operation and classical communication (LOCC):} The concept of local quantum operations and classical communication is a very useful one in quantum information. In an LOCC protocol, a quantum mechanical operation is performed on one of the parts of a  bipartite system and the result of the operation, if it is a measurement, is communicated classically to the other part, where another operation is performed, based on the communicated results. This to and fro communication, interspersed with local quantum operations, may be repeated as many times as needed. The mathematical characterization of LOCC is a difficult one. Recently it was shown that the set of LOCC protocols is not topologically closed but the set of finite-round LOCC is a compact subset of quantum operations \cite{2014CMaPh.328..303C}.
\\
\\
\textbf{Separable operators:} An operation on $\mathcal{L}(\mathcal{B}(H_A),\mathcal{B}(H_B))$ that takes an element $\rho$ $\epsilon$ $\mathcal{L}(\mathcal{B}(H_A),\mathcal{B}(H_B))$ to $\sum_{{i}} (A_i \otimes B_i)\rho (A_i \otimes B_i)^ \dagger$, where $A_i$ and $B_i$ are operations on the Hilbert spaces $H_A$ and $H_B$ respectively, is called separable. An LOCC is certainly a separable operation. The set of separable operations is clearly convex. It is also closed, as shown in the following statement. 
\\
\\
\textbf{Statement:} The set of separable operators is closed in the norm topology.
\\
\\
\textbf{Proof:} It is known that separable operators correspond to separable states under the CJKS isomorphism \cite{2001PhRvL..86..544C}.   The set of separable states is closed and as we have seen in the Lemma above, the CJKS isomorphism is continuous. This implies that the inverse image of the set of separable states under the  CJKS map is closed, i.e., the set of separable operators is closed. \hfill \(\square\)
\\
\\
In the case of constructing a witness for an entangled state, we use the fact that the set of separable states is convex and closed. The witness is constructed under the belief that the corresponding experimental set-up creates that entangled state. The witness, then, detects the state not only if the set-up is perfect but also if there are imperfections present. Notwithstanding any such noise in the set-up, provided the noise is not higher then a certain threshold, the witness guarantees the presence of entanglement in the state created by the set-up. 
\\
\\In a similar way, let $\Lambda$ be a physical operator on density matrices in $\mathcal{L}(\mathcal{B}(H_A),\mathcal{B}(H_B))$ and consider an experimental apparatus that we believe to be implementing the operator. Ideally, we would like to find a witness to detect if the apparatus is implementing a map that is outside the LOCC class. This is a difficult problem. We however use the fact that the closed and convex set of separable physical operations on $\mathcal{L}(\mathcal{B}(H_A),\mathcal{B}(H_B))$ is mapped onto the closed and convex set of separable density matrices on $H_1 \otimes H_2$ and vice-versa, via the CJKS isomorphism, where $H_1=H_A \otimes H_{\tilde{A}}$, $H_2=H_B \otimes H_{\tilde{B}}$, with $\dim H_{\tilde{A}}= \dim H_A$, $\dim H_{\tilde{B}}= \dim H_B$ \cite{2001PhRvL..86..544C}. Therefore, given an apparatus that claims to implement $\Lambda$, we apply $\Lambda$ on the $H_A \otimes H_B$ part of the state (unnormalized) $\sum_{{i}=1}^{\dim H_A} $ $\sum_{{j}
 =1}^{\dim H_B} $ $\ket{ij}_{AB}\ket{ij}_{\tilde{A}\tilde{B}} $, assuming the latter to be available as a resource. The output state, $\rho_{AB\tilde{A}\tilde{B}}$, if non-separable as a density matrix in the $A\tilde{A}:B\tilde{B}$ partition, will imply that the apparatus is implementing an operation that is non-separable in the $A:B$ partition. The non-separability of $\rho_{AB\tilde{A}\tilde{B}}$ in the $A\tilde{A}:B\tilde{B}$ can now be checked by using the concept of entanglement witness for bipartite states. In particular, if  $\rho_{AB\tilde{A}\tilde{B}}^{T_{A\tilde{A}}}$ has a negative eigenvalue, then the partial transpose of the projector of a corresponding eigenvector can act as a witness. Since the LOCC maps are contained within the set of separable operations, $\Lambda$ will of course be non-LOCC. Moreover, the non-LOCC-ness is being detected here not just for $\Lambda$, but for any other operation that is sufficiently close to $\Lambda$, as we have already seen that the set of separable operations forms a closed set. We therefore have a method for witnessing the nonseparability of an arbitrary apparatus that acts on bipartite quantum states, provided the corresponding CJKS state is entangled. Let us now exemplify the method by using an apparatus that claims to implement the CNOT gate for   $\mathbb{C}^2 \otimes \mathbb{C}^2$. 
\\
\\
\textbf{Witness operator for CNOT gate:} The CNOT gate is a paradigmatic non-LOCC operator acting on $\mathbb{C}^2 \otimes \mathbb{C}^2$, that has been implemented in several physical systems. The CJKS state corresponding to the CNOT gate is $\rho_{CNOT}$, being given by
$4\rho_{CNOT}=$
\begin{equation}
\label{noteq}
\left( {\begin{array}{cccccccccccccccc}
1 &	0 &	0 &	0 &	0 &	1 &	0 &	0 &	0 &	0 &	0 &	0 &	0 &	0 &	0 &	0\\
0 &	0 &	0 &	0 &	0 &	0 &	0 &	0 &	0 &	0 &	1 &	0 &	0 &	0 &	0 &	1\\
0 &	0 &	0 &	0 &	0 &	0 &	0 &	0 &	0 &	0 &	0 &	0 &	0 &	0 &	0 &	0\\
0 &	0 &	0 &	0 &	0 &	0 &	0 &	0 &	0 &	0 &	0 &	0 &	0 &	0 &	0 &	0\\
0 &	0 &	0 &	0 &	0 &	0 &	0 &	0 &	0 &	0 &	1 &	0 &	0 &	0 &	0 &	1\\
1 &	0 &	0 &	0 &	0 &	1 &	0 &	0 &	0 &	0 &	0 &	0 &	0 &	0 &	0 &	0\\
0 &	0 &	0 &	0 &	0 &	0 &	0 &	0 &	0 &	0 &	0 &	0 &	0 &	0 &	0 &	0\\
0 &	0 &	0 &	0 &	0 &	0 &	0 &	0 &	0 &	0 &	0 &	0 &	0 &	0 &	0 &	0\\
0 &	0 &	0 &	0 &	0 &	0 &	0 &	0 &	0 &	0 &	0 &	0 &	0 &	0 &	0 &	0\\
0 &	0 &	0 &	0 &	0 &	0 &	0 &	0 &	0 &	0 &	0 &	0 &	0 &	0 &	0 &	0\\
0 &	1 &	0 &	0 &	1 &	0 &	0 &	0 &	0 &	0 &	0 &	0 &	0 &	0 &	0 &	0\\
0 &	0 &	0 &	0 &	0 &	0 &	0 &	0 &	0 &	0 &	0 &	1 &	0 &	0 &	1 &	0\\
0 &	0 &	0 &	0 &	0 &	0 &	0 &	0 &	0 &	0 &	0 &	0 &	0 &	0 &	0 &	0\\
0 &	0 &	0 &	0 &	0 &	0 &	0 &	0 &	0 &	0 &	0 &	0 &	0 &	0 &	0 &	0\\
0 &	0 &	0 &	0 &	0 &	0 &	0 &	0 &	0 &	0 &	0 &	1 &	0 &	0 &	1 &	0\\
0 &	1 &	0 &	0 &	1 &	0 &	0 &	0 &	0 &	0 &	0 &	0 &	0 &	0 &	0 &	0
\end{array}}\right),
\end{equation}
 where we have used the computational basis in $AB\tilde{A}\tilde{B}$ to express the density matrix, and where we have adopted the convention in which the identity superoperator in the CJKS isomorphism is applied onto the second system. $\rho_{CNOT}^{T_{A\tilde{A}}}$ has a single negative eigenvalue and it is non-degenerate. The partial transpose in the $A\tilde{A}:B\tilde{B}$ partition of the projection on the corresponding eigenvector is given by
\\ 
\begin{equation}
W=\left(
\begin{array}{cccccccccccccccc}
 0 & 0 & 0 & 0  & 0 & 0 & 0 & 0  & 0 & 0 & 0& -1  & 0 & 0 & -1 &0 \\
 0 & 1 & 0 & 0  & 1 & 0 & 0 & 0  & 0 & 0 & 0 & 0  & 0 & 0 & 0 & 0 \\
 0 & 0 & 0 & 0  & 0 & 0 & 0 & 0  & 0 & 0 & 0 & 0  & 0 & 0 & 0 & 0 \\
 0 & 0 & 0 & 0  & 0 & 0 & 0 & 0  & 0 & 0 & 0 & 0  & 0 & 0 & 0 & 0 \\
 
 0 & 1 & 0 & 0  & 1 & 0 & 0 & 0  & 0 & 0 & 0 & 0  & 0 & 0 & 0 & 0 \\
 0 & 0 & 0 & 0  & 0 & 0 & 0 & 0  & 0 & 0 & 0& -1  & 0 & 0 & -1 &0 \\
 0 & 0 & 0 & 0  & 0 & 0 & 0 & 0  & 0 & 0 & 0 & 0  & 0 & 0 & 0 & 0 \\
 0 & 0 & 0 & 0  & 0 & 0 & 0 & 0  & 0 & 0 & 0 & 0  & 0 & 0 & 0 & 0 \\
 
 0 & 0 & 0 & 0  & 0 & 0 & 0 & 0  & 0 & 0 & 0 & 0  & 0 & 0 & 0 & 0 \\
 0 & 0 & 0 & 0  & 0 & 0 & 0 & 0  & 0 & 0 & 0 & 0  & 0 & 0 & 0 & 0 \\
 0 & 0 & 0 & 0  & 0 & 0 & 0 & 0  & 0 & 0 & 1 & 0  & 0 & 0 & 0 & 1 \\
 -1& 0 & 0 & 0  & 0 & -1& 0 & 0  & 0 & 0 & 0 & 0  & 0 & 0 & 0 & 0 \\
 
 0 & 0 & 0 & 0  & 0 & 0 & 0 & 0  & 0 & 0 & 0 & 0  & 0 & 0 & 0 & 0 \\
 0 & 0 & 0 & 0  & 0 & 0 & 0 & 0  & 0 & 0 & 0 & 0  & 0 & 0 & 0 & 0 \\
 -1& 0 & 0 & 0  & 0 & -1& 0 & 0  & 0 & 0 & 0 & 0  & 0 & 0 & 0 & 0 \\
 0 & 0 & 0 & 0  & 0 & 0 & 0 & 0  & 0 & 0 & 1 & 0  & 0 & 0 & 0 & 1 \\
\end{array}
\right),
\end{equation}
 where again we have used the computational basis in $AB\tilde{A}\tilde{B}$ to express the matrix. This operator can now act as a witness operator for detecting the non-separability of $\rho_{CNOT}$ in the $A\tilde{A}:B\tilde{B}$ partition, thereby implying the non-separability (i.e., being not a separable operator) of $\Lambda$ on $H_A \otimes H_B$. This in turn will imply that $\Lambda$ is not an LOCC on  $H_A \otimes H_B$.

The decomposition of the witness operator $W$ is given by
\begin{align}
W=\sum_{{i,j,k,l}=1}^{4} w_{ijkl}(-1)^\alpha \mu_i \otimes \mu_j \otimes \mu_k \otimes \mu_l ,
\end{align}
where $\mu_1=I+\sigma_z $, $\mu_2=I-\sigma_z $, $\mu_3=\sigma_x+i\sigma_y $,$\mu_4=\sigma_x-i\sigma_y $.
$\alpha$ depends on $i,j,k,l$, and is equal to the total number of appearances of the indices 3 and 4 in the corresponding term in the summation. There are only 16 non-zero terms in the summation, viz. for \begin{eqnarray}
&&\nonumber(i,j,k,l)=\\
&&\nonumber(1,1,1,2),(1,4,1,3),(4,1,4,4),(4,4,4,1),\\
&&\nonumber(1,3,1,4),(1,2,1,1),(4,3,4,2),(4,2,4,3),\\
&&\nonumber(3,1,3,3),(3,4,3,2),(2,1,2,1),(2,4,2,4),\\
&&\nonumber(3,3,3,1),(3,2,3,4),(2,3,2,3),(2,2,2,2), 
\end{eqnarray}
and $w_{ijkl}=1$ for these combinations, being vanishing for others. Since W has been decomposed into a sum over local operations on $H_A \otimes H_{\tilde{A}} \otimes H_B \otimes H_{\tilde{B}}$, we have a local strategy for measuring $W$.
\\
\\
\textbf{Noise analysis:} We now try to quantify the amount of noise that can be inflicted on the CNOT gate, and yet its nonseparability can be detected by the witness. For this purpose, we use the completely depolarizing noise, i.e., for a given map $\Lambda$, we consider the noisy map, $p\Lambda + (1-p)D$, where $D$ is the completely depolarizing map, and $0 \leq p \leq 1$. By considering the CJKS state corresponding to the noisy map, we found that the nonseparability is still detected provided $p > \frac{1}{9}$.
\\
\\
\textbf{Comparison with situation when resource is absent:} The nonseparability of CNOT gate can of course be detected even when the maximally entangled state in $AB:\tilde{A}\tilde{B}$ is unavailable. In this case, we can try to detect the nonseparability of the CNOT gate by applying it on $\ket{+}_A \ket{0}_B$, whereby a maximally entangled two-qubit state is created, where $\ket{+}$ and $\ket{0}$ are respectively eigenstates of $\sigma_x$ and $\sigma_z$ corresponding to the eigenvalues $+1$ for each. The nonseparability is then detected by witnessing the entanglement in the output state. To compare this method with the case when the CJKS map is employed, we again consider the noisy CNOT map, $p\Lambda_{CNOT} + (1-p)D$, apply it on $\ket{+}_A \ket{0}_B$, and try to witness the entanglement in the output state. In this case, the nonseparability can be detected provided $p> \frac{1}{3}$. We therefore find that there is a clear trade-off between the efficiency of the nonseparability detection, as quantified by the amount of noise that the detection process can withstand, and the resource available. Provided that a resource, in the form of a maximally entangled state in $AB:\tilde{A}\tilde{B}$, is available, the noise tolerance is much higher.
\\
\\
\noindent \textbf{Witness operator for the square root of  swap gate:} We now repeat the above procedure for the square root of the swap gate, where the swap gate is defined as a two-party gate (whose two-qubit version we are concerned with here) that interchanges the states of any product input, with the action on other states being defined linearly. The witness operator for this gate, corresponding to the negative eigenvalue,  $-\frac{\sqrt{5}}{8}$, of the partial transpose of the CJKS state of this gate is given by 
%
\begin{align}
W_{\sqrt{swap}}=\sum_{{i,j,k,l}=1}^{4} w^s_{ijkl} \mu_i \otimes \mu_j \otimes \mu_k \otimes \mu_l ,
\end{align} 
where $w^s_{ijkl}=1$ for 
\begin{eqnarray}
&&(i,j,k,l)=\nonumber\\
&&(1,1,1,2),(1,1,2,1),(1,1,2,2),(3,4,1,2),\nonumber
\\
&&(3,4,2,1),(1,2,1,1),(1,2,3,4),(1,2,4,3),\nonumber \\
&&(1,2,2,2),(2,1,1,1),(2,1,3,4),(2,1,4,3),\nonumber \\
&&(2,1,2,2),(4,3,1,2),(4,3,2,1),(2,2,1,1),\nonumber \\&&(2,2,1,2),
(2,2,2,1), \nonumber 
\end{eqnarray}
 $w^s_{ijkl}=-1$ for 
 \begin{eqnarray}
 &&\nonumber (i,j,k,l)=\\
 &&\nonumber (1,4,4,1),(1,4,2,4),(4,2,4,1),(4,2,2,4),\\
 && \nonumber (3,1,1,3),(3,1,3,2),(2,3,1,3),(2,3,3,2),\\
 && \nonumber (3,4,4,3),(1,3,3,1),(1,3,2,3),(3,2,3,1),\\
 && \nonumber (3,2,2,3),(4,1,1,4),(4,1,4,2),(2,4,1,4),\\&& \nonumber (2,4,4,2),(4,3,3,4) ,
 \end{eqnarray}
 and $w^s_{ijkl}=0$ for the 
 remaining combinations  of $(i,j,k,l)$.
 \\
 \\
 \noindent \textbf{Noise analysis for the square root of swap gate and comparison with resource-free case:}
 Just like for the CNOT gate, we again perform a noise analysis here by considering admixture with the completely depolarizing channel. We find that when the maximally entangled state is available as a resource, the nonseparability in the square of the swap gate can be detected 
 when $p>\frac{1}{1+2\sqrt{5}} \approx 0.18$. For the case when the resource is not available, the nonseparability can be detected 
 when $p>\frac{1}{3}$, corresponding to the situation when the square root of the swap gate is applied to the state \(|01\rangle\), which produces a maximally entangled two-qubit state.
 \\
 \\
 \textbf{Witness operator for the Bell gate:} Let us now consider yet another example of an entangling gate, viz.   the Bell gate \cite{nahoi-mon-ditei-tumi-parona, ei-akash-natun}, defined as one that respectively takes the states of the standard biorthogonal two-qubit product basis consisting of the states \(|00\rangle\), \(|11\rangle\), \(|01\rangle\), \(|10\rangle\),
 to those of the Bell basis consisting of the states \(|00\rangle + |11\rangle)/\sqrt{2}\), \(|00\rangle - |11\rangle)/\sqrt{2}\), 
 \(|01\rangle + |10\rangle)/\sqrt{2}\), \(|01\rangle - |10\rangle)/\sqrt{2}\).
 %
 The decomposition of the witness operator corresponding to the eigenvalue $-\frac{1}{2}$ of the partial transpose of the CJKS state corresponding to this gate is given by 
 \begin{align}
W_{Bell}=\sum_{{i,j,k,l}=1}^{4} w^B_{ijkl} \mu_i \otimes \mu_j \otimes \mu_k \otimes \mu_l ,
\end{align} \\
where $w^B_{ijkl}=1$ for 
\begin{eqnarray}
&&\nonumber(i,j,k,l)=\\
&&\nonumber(1,1,1,2),(1,1,2,1),(1,4,1,3),(1,4,2,4),\\
&&\nonumber(3,1,4,1),(3,1,2,4),(3,4,4,4),(3,4,2,1),\\
&&\nonumber(1,3,1,4),(1,3,2,3),(1,2,1,1),(1,2,2,2),\\
&&\nonumber(3,3,4,3),(3,3,2,2),(3,2,4,2),(3,2,2,3),\\
&&\nonumber(4,1,3,1),(4,1,2,3),(4,4,3,4),(4,4,2,2),\\
&&\nonumber(2,1,1,1),(2,1,3,4),(2,1,4,3),(2,1,2,2),\\
&&\nonumber(2,4,1,4),(2,4,3,1),(2,4,4,2),(2,4,2,2),\\
&&\nonumber(4,3,3,3),(4,3,2,1),(4,2,3,2),(4,2,2,4),\\
&&\nonumber(2,3,1,3),(2,3,3,2),(2,3,4,1),(2,3,2,4),\\
&&\nonumber(2,2,1,2),(2,2,3,3),(2,2,4,4),(2,2,2,1), 
\end{eqnarray}
$w^B_{ijkl}=-1$ for
\begin{eqnarray}
&&\nonumber(i,j,k,l)=\\
&&\nonumber(1,1,3,3),(1,1,4,4),(1,4,3,2),(1,4,4,1),\\
&&\nonumber(3,1,1,3),(3,1,3,2),(3,4,1,2),(3,4,3,3),\\
&&\nonumber(1,3,3,1),(1,3,4,2),(1,2,3,4),(1,2,4,3),\\
&&\nonumber(3,3,1,1),(3,3,3,4),(3,2,1,4),(3,2,3,1),\\
&&\nonumber(4,1,1,4),(4,1,4,2),(4,4,1,1),(4,4,4,3),\\
&&\nonumber(4,3,1,2),(4,3,4,4),(4,2,1,3),(4,2,4,1), 
\end{eqnarray}
and $w^B_{ijkl}=0$ for the rest of the different values of $(i,j,k,l)$.
\\
\\
\noindent \textbf{Noise analysis for the Bell gate and comparison with resource-free case:}
 In this case, the nonseparability is detected via the CJKS approach when $p>\frac{1}{9}$ in the noisy case, and it is detected without the resource when $p>\frac{1}{3}$, in the situation where the Bell gate is applied on \(|00\rangle\).
\section{Conclusion}
\label{karane-akaran}
The concept of entanglement witness forms an efficient method for detecting entangled states. Entangled states are created by quantum processes that cannot be written in separable form. We show that it is possible to conceptualize a parallel method of witness for nonseparability of quantum processes. The method utilizes the channel-state duality between quantum channels and states. The method uses a maximally entangled state as a resource. We find that considering the witness for noisy maps provide us with a trade-off between the threshold amount of noise that the method can withstand and the entanglement in the resource.
\section*{Acknowledgments}
We acknowledge useful discussions with Aditi Sen(De).

\bibliographystyle{apsrev4-1}
\bibliography{Reference}

\end{document}